\documentclass[fleqn,usenatbib]{mnras}

\usepackage[T1]{fontenc}

\DeclareRobustCommand{\VAN}[3]{#2}
\let\VANthebibliography\thebibliography
\def\thebibliography{\DeclareRobustCommand{\VAN}[3]{##3}\VANthebibliography}

\usepackage{graphicx}	
\usepackage{amsmath}	
\usepackage{amssymb}	

\def\h2o{H$_2$O}
\def\nh3{NH$_3$}

\def\12CO{$^{12}$CO}
\def\13CO{$^{13}$CO}
\def\C18O{C$^{18}$O}
\def\CH3OH{CH$_3$OH}
\def\g22{G22.357+0.066}
\def\g25{G25.411+0.105}
\def\g9{G9.62+0.19E}

\title[Orientation Effect on the Light Curve Shape of Periodic Methanol Maser
  Sources]{Orientation Effect on the Light Curve Shape of Periodic Methanol Maser Sources}

\author[J. Morgan et al.]{
J. Morgan,$^{1}$\thanks{E-mail:jeanmariemorgan0@gmail.com (JM)}
D.J van der Walt,$^{1}$
J.O. Chibueze$^{1,2}$
and Q. Zhang$^{3}$
\\
$^{1}$Centre for Space Research, North-West University,
Potchefstroom 2520, South Africa\\
$^{2}$Department of Physics and Astronomy, Faculty of Physical Sciences,  University of Nigeria, \\Carver Building, 1 University Road,
Nsukka, Nigeria\\
$^{3}$Center for Astrophysics $|$ Harvard \& Smithsonian , 60 Garden Street, Cambridge MA 02138, USA }
\date{Accepted XXX. Received YYY; in original form ZZZ}

\pubyear{2020}

\begin{document}
\label{firstpage}
\pagerange{\pageref{firstpage}--\pageref{lastpage}}
\maketitle

\begin{abstract}
We report the results of our pilot millimeter observations of periodic maser sources. Using SMA, we carried out 1.3\,mm observations of G22.357$+$0.066 and G25.411$+$0.105, while ALMA 1.3 mm archival data was used in the case of G9.62+0.19E. Two continuum cores (MM1 and MM2) were detected in G22.357$+$0.066, while 3 cores (MM1 -- MM3) detected in G25.411$+$0.105. Assuming dust-to-gas ratio of 100, we derived the masses of the detected cores. Using the $^{12}$CO (2-1) and $^{13}$CO (2-1) line emission, we observed gas kinematics tracing the presence of bipolar outflows in all three star-forming regions. In the cases of G22.357$+$0.066 and G9.62+0.19E, both with similar periodic maser light curve profiles, the outflowing gas is seen in the north-west south-east direction. This suggest edge-on view of the 2 sources. G25.411$+$0.105, with a contrasting light curve profile, show a spatially collocated blue and red outflow lobes, suggesting a face-on view. Our results suggest that orientation effects may play a role in determining the characteristics of the light curves of periodic methanol masers.
\end{abstract}




\section{Introduction}
Limitation in the number of samples of massive stars (compared to low-mass counterparts), 
their generally large distances from the Sun and along with
being deeply embedded in their natal clouds for most of their lifetimes, impede our
understanding of various evolutionary stages of massive stars. The challenges of the
observational study of high-mass star formation are gradually being tackled with improved
sensitivities and angular resolutions offered by new and upgraded instruments around the
globe. Some of the recently resolved issues include the problem of angular momentum
transfer to enable mass accretion onto the central core (young protostar). The Atacama
Large Millimetre/sub-millimetre Array (ALMA) offered ideal data for detecting a
disk-driven rotating bipolar outflow in Orion Source I \citep{Hirota2017}. A keplerian
disk has also been detected in the massive young stellar object G11.92$-$0.61 MM1
\citep{Ilee2016}.

Masers are known to serve as signposts of sites for star formation, especially the 6.7 GHz
class II methanol (CH$_3$OH) masers that are known to be exclusively associated with
massive stars \citep{Breen2013}. They could serve as tracers of physical conditions around forming high-mass stars. The variability of CH$_3$OH masers was first outlined by
\citet{Goedhart2004} who discovered periodicity in the flux density variations of class II
CH$_3$OH masers. Since then, an additional 25 periodic class II methanol masers, with periods ranging between 24 and 510 days, have been discovered. An intriguing property of
the sample of known periodic methanol masers is that there is no single unique shape of
the light curves of these masers. It is not possible to classify all periodic masers
according to the shape of their light curves, however, \citet{Maswanganye2015} suggests that there are at least two categories of periodic  masers for which the light curves are rather well defined: (i) The flare profile shows a fast increase in intensity from a low state followed by a slow decrease in intensity which can be described in terms of the recombination of a hydrogen plasma from a higher to a lower state of ionization.  The sources belonging to this group are G9.62+0.19E \citep{Goedhart2003}, G22.357$+$0.066 \citep{Szymczak2015}, G37.55+0.20 \citep{Araya2010}, and G45.473+0.134 \citep{Szymczak2015}.  (ii) The variability profile resembles that of $\left|\cos(\omega t)\right|$ function, (referred to as a Bunny hop) and is similar to that seen in some eclipsing binary systems. Sources belonging to this group are G25.411+0.105 \citep{Szymczak2015}, G339.62$-$0.12 \citep{Goedhart2014}, G338.93$-$0.06
\citep{Goedhart2014} and G358.460$-$0.391 \citep{Maswanganye2015}.

Several mechanisms have been proposed to explain the periodicity of methanol
masers. \citet{Vanderwalt2009} and \citet{Vanderwalt2011} proposed a colliding wind binary (CWB) system for the periodic methanol masers in G9.62+0.20E and other periodic methanol maser sources with similar light curves. Using a more realistic model \citet{Vandenheever2019} also showed that the CWB scenario can explain the light curves of the periodic methanol masers in G9.62+0.20E, G22.357$+$0.066, G37.55+0.20, and G45.473+0.134. Other mechanisms proposed are: the pulsational instabilities of very young accreting high-mass stars \citep{Inayoshi2013}, spiral shocks
associated with very young binary systems orbiting within a circumbinary disk \citep{Parfenov2014}, periodic accretion in a very young binary system \citep{Araya2010}, and outflows in a binary system \citep{Singh2012}. \citet{Maswanganye2015} evoked an eclipsing binary system involving a high-mass (primary) star and a bloated low-mass (companion) star in explaining the observed periodicity of the methanol masers in  G358.460-0.391.
While each of the suggested mechanisms may be sufficient for explaining periodicity in
individual star-forming regions, the complexity in the nature of massive star-forming environments and the uniqueness of each region, make it difficult to evoke any one of the mechanisms for all cases. In fact, different mechanisms may be responsible for the different observed flare profiles \citep[see][for a discussion]{Vanderwalt2016}. However, the fact that it is possible to identify at least two groups of periodic sources (as explained above), each for which the light curves are very similar, suggests that there are some sources for which the underlying mechanism is the same. It is therefore reasonable to argue that there might be other similarities in star-forming regions that host periodic methanol masers with same light curves. In other words, the similarity in light curves might also manifest in other properties of the star-forming regions in addition to the maser emission.

To understand the cause of periodicity in class II CH$_3$OH masers, it is imperative to
probe the bulk of the gas and dust beyond the masing region which constitutes only a small
fraction of the star-forming environment. Observing dust continuum and molecular line
emissions of periodic maser sources could reveal some physical properties that are key to
understanding periodicity in class II CH$_3$OH maser sources.

In this paper, we present the millimetre properties of three periodic CH$_3$OH maser sources viz., G22.357+0.066 [kinematic distance 4.86 kpc based on \citet{Reid2009}; see also
  \citep{Szymczak2011}], G25.411+0.105[kinematic distance of 9.50 kpc   \citep{Szymczak2007}, while \citet{Sridharan2002} reported the presence of SiO,
  CH$_3$OH, and CH$_3$CN molecular lines as well as H$ _{2} $O and CH$_3$OH masers.] and
G9.62+0.19E[parallax distance of 5.2 kpc \citep{Sanna2009}. \citet{Sanna2009} also
  reported the presence of a 1.3 cm continuum source detected with VLA.]. G9.62+0.19E and
G22.357$+$0.066 have the same type of flaring light curve while G25.411+0.105's light curve
is of the Bunny hop type.
\section{Observations}
\label{sec:obs}
\subsection{Submillimeter Array (SMA) Observations}
 The SMA\footnote{The Submillimeter Array is a joint project between the Smithsonian
  Astrophysical Observatory and the Academia Sinica Institute of Astronomy and
  Astrophysics, and is funded by the Smithsonian Institution and the Academia Sinica.}
observations of G22.357$+$0.066 and G25.411+0.105 were carried out on June 27, 2016, and August 6, 2017.  The phase
center for G22.357+0.066 and G25.411+0.105 is (R.A., decl.)$ _{J2000} $ $ = $ (18$
^{h} $ 31$ ^{m} $ 44$ ^{s} $.29, -09$ ^{d} $ 22$ ^{m} $ 11$ ^{s} $.2) and (R.A., decl.)$
_{J2000} $ $ = $ (18$ ^{h} $ 37$ ^{m} $ 16$ ^{s} $.90, -06$ ^{d} $ 38$ ^{m} $ 30$ ^{s}
$.5), respectively with a 2$\prime\prime$ error in the pointing.

The observations in 2016 used the ASIC and SWARM correlator with seven antennas in the compact array configuration.  The 8 GHz bandwidth is divided into two side-bands, lower side-band of 4 GHz and upper side-band of 4 GHz, each consisting of 128 channels. The lower side-band covered the frequency range from 212.9 GHz to 220.6 GHz and the upper side-band the frequency range from 228.7 GHz to 236.6 GHz. The calibrators and the two target sources were observed in turn, leading to a total observation time of 80 minutes for each target source. The system temperature correction was done in MIRIAD and the data was converted to CASA measurements sets. The full calibration of the data was done in CASA, with Titan as flux calibrator and 3C279 as bandpass calibrator. The antenna gain correction was done using Quasars J1733$-$130 and J1743$-$038.

The observations on August 6, 2017 used six antennas in the compact array configuration. The SWARM correlator was configured to cover 8 GHz bandwidth in lower side-band (213.5 $-$ 221.5 GHz) and upper side-band (229.5 $-$ 237.5 GHz), respectively. The native spectral resolution is 140 kHz per channel across the entire spectral band. The on-source time for each target is approximately 60 min. Neptune was observed for flux calibration, quasar 3C279 for bandpass calibration and quasars J1830$+$063 and J1743$-$038 for time-dependent antenna gain calibration. The raw visibilities were calibrated using the IDL superset 
MIR\footnote{https://www.cfa.harvard.edu/$\sim$cqi/mircook.html}, and exported into CASA for imaging.

The 2016 and 2017 data sets were combined in CASA and the largest recoverable scale is $\sim$ 33 $^{\prime\prime}$. The CASA clean task was used to produce the molecular line emission spectra of $^{12}$CO, $^{13}$CO and C$^{18}$O with their rest frequencies at 230.53800000 GHz, 220.39868420 GHz and 219.56035410 GHz respectively. The rest frequencies were obtained from the Splatalogue catalogue; more details on the line emission parameters is given in Table \ref{tab:line parameters}. The channel width for all the line emission is 1.0 km s$ ^{-1} $.  Other than the CO lines, there were no significant line features above 3$\sigma$. The dust continuum were produced using the line-free channels. The synthesized beam size for the continuum is 6$^{\prime\prime}$.08 $\times$ 2 $^{\prime\prime}$.79 and 5$^{\prime\prime}$ .80 $\times$ 2$^{\prime\prime}$ .50 for the spectral line images.The root-mean-square (rms) noise level in the continuum, $^{12}$CO line images and  $^{13}$CO/C$^{18}$O line images are $\sim$ 1 mJy beam$^{-1}$, 170 mJy\,beam$^{-1}$ and $\sim$ 20 mJy\,beam$^{-1}$ respectively.

\subsection{ALMA Archival Data}
We obtained ALMA band 6 archival data on G9.62+0.19 (Project ID: 2013.1.00957.S) taken
on 2015 April 27 (39 antennas of the 12-m main array), May 24 (34 antennas of the 12-m
main array) in its compact and extended configurations, respectively. It was also observed
with the Atacama Compact Array (ACA: 7-m array) on 2015 May 04 with 10
antennas. J1733$-$1304, J1517$-$2422 were used as bandpass calibration and phase
calibration, respectively, while Neptune, Titan and J1733$-$130 were used as flux calibrators. 

 The Band 6 (230 GHz) receivers were used for the observations in dual-polarization mode. To cover the molecular lines  $ ^{12} $CO (2-1) and CH$_{3}$OH  {\bf $v_{t} = 1$, (6$_{1}$-7$_{2}$ A-)}, five spectral windows with a bandwidth of 117 MHz in each window, as well as another spectral window with a bandwidth of 234 MHz, was used. The channel width in all the spectra windows is 0.08 km s$^{-1}$. The ALMA data was also calibrated in CASA.  They combined data from three configurations in CASA, that included single dish data, with a LAS of $\sim$ 20 $^{\prime\prime}$ for the combined data.

The continuum image has a rms noise of 1.0 mJy\,beam$^{-1}$ and a synthesized beam of 0$^{\prime\prime}$.94 $\times$ 0 $^{\prime\prime}$.71. For the molecular lines the rms noise is $\sim$ 20 mJy\,beam$^{-1}$. The lines $^{12}$CO (2-1) and CH$_{3}$OH {\bf $v_{t} = 1$, (6$_{1}$-7$_{2}$ A-)} has synthesized beam sizes of 0$^{\prime\prime}$.94 $\times$ 0 $^{\prime\prime}$.73 and 0$^{\prime\prime}$.99 $\times$ 0 $^{\prime\prime}$.72 receptively.  The rest frequency used for CH$_{3}$OH {\bf $v_{t} = 1$, (6$_{1}$-7$_{2}$ A-)} was 217.29920 GHz and was also obtained from the Splatalogue; the details on the line parameters can be found in Table \ref{tab:line parameters}. More details of the observations can be found in \citet{Liu2017}.
\section{Results} 
\label{sec:results}

\subsection{G22.357$+$0.066} 
\subsubsection{1.3~mm Continuum Emission: G22.357$+$0.066}
\label{subsec:G22}

\begin{figure}
	\includegraphics[width=\columnwidth]{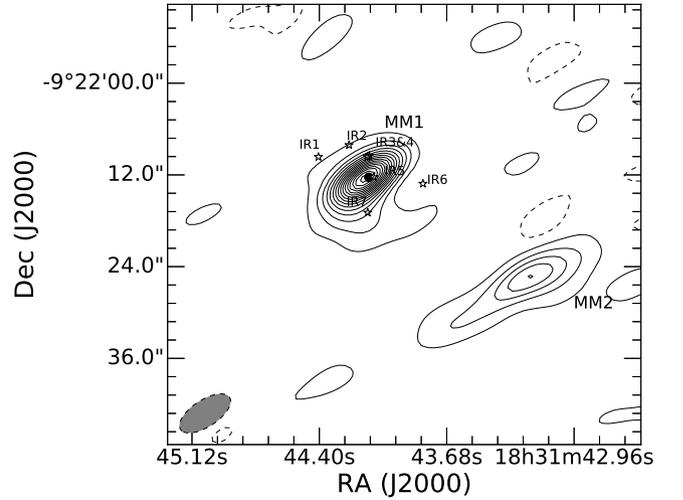}
    \caption{ G22.357$+$0.066: 1.3 $mm$ continuum emission with the near infra-red (NIR)
      counterparts (star symbol) obtained from \citet{UKIDSS2012} within a radius of 5 $
      ^{\prime \prime} $ from the maser. MM1 and MM2 indicates each millimetre object
      detected in the SMA observation. The dust contour levels are [-3, 3, 9, 15, 21, 27, 33, 39, 45, 51, 57, 63, 69, 75, 81, 87, 93] $\times$ $\sigma$ mJy\,beam$^{-1}$, where $\sigma$ is 1mJy\,beam$^{-1}$ and the circle shows the position of the periodic methanol maser which has a positional uncertainty of $\sim$ 6$^{\prime\prime}$.}
    \label{fig:G22_continuum}
\end{figure}

Two millimetre continuum cores (MM1 \& MM2) associated with G22.357$+$0.066, which is located at a distance 4.86 $\pm$ 0.3 kpc \citep{Reid2009}, were detected in the SMA observations and are presented in Figure \ref{fig:G22_continuum}.The root-mean-square (rms) noise level in the continuum image of G22.357+0.066 is  $\sim$ 1 {\bf mJy\,beam$^{-1}$}. Table \ref{tab:cont parameters} presents the peak positions (RA and Dec), the peak and integrated flux densities for each detected continuum object. MM1 is the only resolved core. MM2 is marginally resolved with an angular size of 6$^{\prime\prime}$.8 $\times$ 3$^{\prime\prime}$.1. MM2 is located $ \leq $ 16$^{\prime\prime}$, south-west of MM1. Using the methanol multi-beam (MMB) survey position of the masers, we found it to be located $\sim$\,0\arcsec.2 from the peak continuum emission of MM1 (see Figure \ref{fig:G22_continuum}).

\begin{table*}
	\centering
	\caption{Continuum position parameters}
	\label{tab:cont parameters}
	\begin{tabular}{lccccr} 
		\hline
		Object-name & R.A. & Dec. & Peak flux & Integrated flux & P.A \\ 
            & (h m s) & ($ ^{\circ}$~$\prime$~$\prime\prime$) & (mJy\,beam$^{-1}$) & (mJy) &  deg\\ 
		\hline
G22.357$+$0.066 - MM1 &  18 31 44.15 &  -09\ 22\ 12.6 & 92.1 &  94.5 &  -37.1\\
G22.357$+$0.066 - MM2 &  18 31 43.21 &  -09\ 22\ 25.3 &  22.6 &  27.9 & \\ 

G25.441$+$0.105 - MM1 &  18 37 16.94 &  -06\ 38\ 30.1 & 64.6 & 66.7 &  -36.7\\
G25.441$+$0.105 - MM2 &  18 37 16.56 &  -06\ 38\ 32.4 & 13.7 & 27.2 &\\
G25.441$+$0.105 - MM3 &  18 37 17.18 &  -06\ 38\ 33.4 & 7.1 & 13.9 & \\ 

G9.62+0.19E (MM4) &  18 06 14.66 &  -20\ 31\ 31.6 & 119.1 & 136.5 & 81.9\\ 
 
		\hline
	\end{tabular}
\end{table*}

The dust mass, $M_{d}$, can be estimated from
\begin{equation}
M_{d} = \dfrac{S_{\nu} \, D^{2}}{\kappa_{\nu} \, B_{\nu}(T_{d})} \label{mass equation} 
\end{equation}
assuming optically thin dust emission \citep{Hildebrand1983}. $S_{\nu}$ is the continuum
flux density at frequency, $\nu$, $B_{\nu}(T_{d}) $ is the Planck function at dust
temperature, $T_{d}$, $D$ is the distance to the source and $\kappa_{\nu}$ the dust opacity
per unit mass, $\kappa_{\nu}$ $ =$ 0.19 cm$^{2}$ g$^{-1}$ at 230 GHz
\citep{Draine2001}. Observations indicate that star-forming regions that have no signs of an \ion{H}{II}  regions, but that are associated with masers have temperatures of $\sim$ 20 K \citep{Wu2006, Urquhart2011, Lu2014}. Thus we adopted a dust temperature, $T_{d}$ , of 20 K with a $\pm$ 5 K uncertainty and used a gas-to-dust mass ratio of
100 to estimate the mass of the cores (MM1 \& MM2). The main uncertainty in the
estimation of the core mass is from the $\pm$5 K uncertainty in the dust temperature and the $\pm$0.3 kpc uncertainty in the distance. The mass of the main core (MM1) associated with G22.357$+$0.066 was estimated to be $ \sim $ 175$\pm$20 M$_{\odot}$ and for MM2, assuming the same distance as for MM1, the mass was estimated to be $ \sim $ 52$\pm$6 M$_{\odot}$.

\subsubsection{Carbon-monoxide Emission in G22.357+0.066}
\begin{figure*}
	\includegraphics[width=0.8\textwidth ]{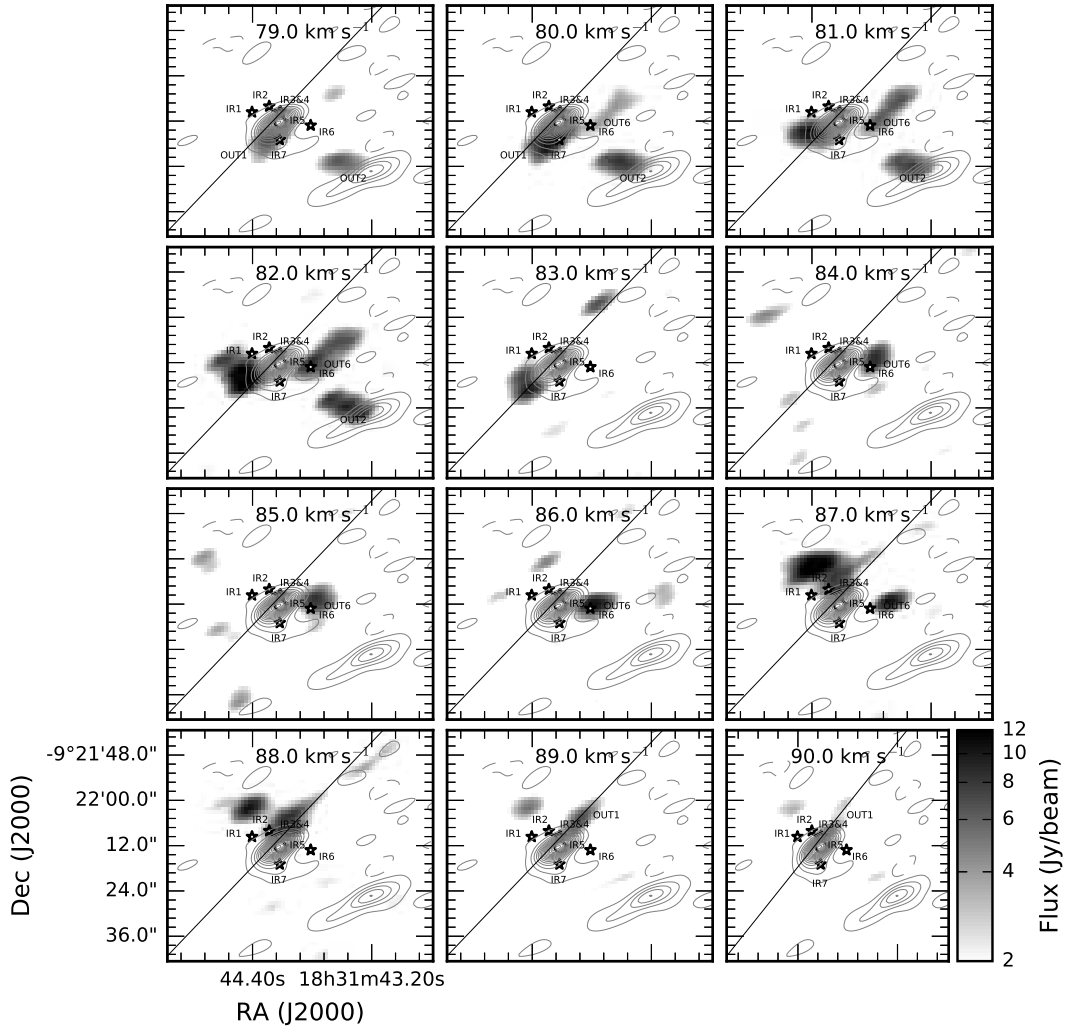}
    \caption{G22.357+0.066: $^{12}$CO line emission velocity channel maps form  79.0
      km s$^{-1}$ to 90.0 km s$^{-1}$.} The position of the periodic methanol
      maser is indicated with a circle and the grey contours represent the dust
      continuum with the same contour levels as indicated in the continuum image. The central velocity of each channel is indicated in the upper panel. The NIR counterparts obtained from the UKIDSS database are indicated with the star symbol. The emission is indicated by the grey scale on the bottom right of the channel maps.
    \label{fig:G22_12CO_chanmaps}
\end{figure*}

\begin{table} 
\setlength{\tabcolsep}{5pt}
	\centering
	\caption{Line parameters}
	\label{tab:line parameters}
	\begin{tabular}{lccc} 
		\hline
		 Molecule & Transition & Rest frequency  & E$_{L}$ \\
         &  & (GHz) & (K) \\
		\hline
  $^{12}$CO  & 2-1             & 230.53800000  & 5.53  \\		
  $^{13}$CO  & 2-1             & 220.39868420  & 5.29  \\
  C$^{18}$O  & 2-1             & 219.56035410  & 5.27  \\
  CH$_{3}$OH & 6$_{1}$-7$_{2}$ A- & 217.29920500 & 363.50 \\
  		\hline
	\end{tabular}
\end{table}
Figure \ref{fig:G22_12CO_chanmaps} presents the channel maps of $^{12}$CO in G22.357$+$0.066. The dust continuum (white contours) and the near infra-red (NIR 1 to 7) counterparts are superimposed on all the channel maps. The position of the periodic 6.7 GHz methanol maser is indicated by a circle. Using the peak velocity of the C$^{18}$O line, \citet{Szymczak2007} determined the systemic velocity of G22.357+0.066 to be at 84.2 km s$^{-1}$. The velocity range of the $^{12}$CO is from 72 km s$^{-1}$ to 91 km s$^{-1}$,  however the channel maps only show the emission from 79.0 km s$^{-1}$ to 90.0 km s$^{-1}$. From 79 km s$^{-1}$ to 82 km s$^{-1}$ there is $^{12}$CO emission (OUT2) associated with MM2 and in channel maps  80 km s$^{-1}$ to 82 km s$^{-1}$  and  84 km s$^{-1}$ to 87 km s$^{-1}$, $^{12}$CO emission (OUT6) is observed that is associated with the near-infrared source NIR6. In channel map 79 km s$^{-1}$ to 80 km s$^{-1}$ $^{12}$CO emission (OUT1) is observed south-east of the dust continuum MM1 and in channel maps 89 km s$^{-1}$ to 90 km s$^{-1}$, $^{12}$CO emission (OUT1) is observed north-west of MM1. The south-east emission is red-shifted and the north-west emission is blue-shifted. This symmetry (south-east - north-west) seen in the emission with regards to the dust continuum (indicated by the line in Figure \ref{fig:G22_12CO_chanmaps}) is interpreted to be tracing a bipolar outflow as discussed in section \ref{Bipolar outflows}.

\subsection{G25.411+0.105} 
\subsubsection{1.3~mm Continuum Emission:G25.411+0.105 }
\label{subsec:G25}

\begin{figure}
	\includegraphics[width=\columnwidth]{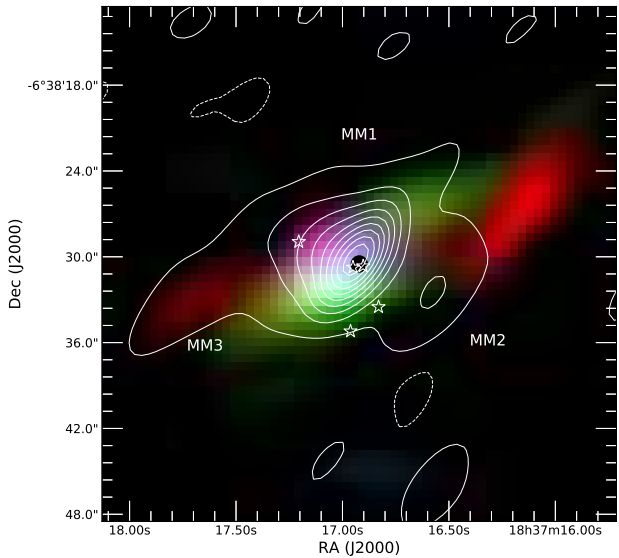}
    \caption{ G25.411+0.105: 3-color image of the CO emission with the $^{12}$CO in red, $^{13}$CO in green and C$^{18}$O in blue.  The 1.3 $mm$ continuum emission contours are overlaid on the CO emission with the NIR counterparts (star symbol) obtained from \citet{UKIDSS2012} within a radius of 5 $ ^{\prime \prime}$ from the maser. MM1 to MM3 indicates each millimetre object detected in the SMA observation. The dust emission contour lines are [-2, 2,  10, 18, 26, 34, 42, 50, 58, 66] $\times$ $\sigma$ mJy\,beam$^{-1}$, with  $\sigma$ as 1 mJy\,beam$^{-1}$. The circle shows the position of the periodic 6.7 GHz methanol maser with a $\sim$ 6$^{\prime\prime}$ uncertainty.}
      \label{fig:G25_continuum} 
 \end{figure}

Figure \ref{fig:G25_continuum} shows the millimetre continuum objects (MM1 to MM3) detected in G25.411+0.105, with the rms noise level  $\sim$ 1 {\bf mJy\,beam$^{-1}$}. Table \ref{tab:cont parameters} presents the peak positions (RA and Dec), the peak and integrated flux densities for each detected continuum object. MM2, located west of MM1, has an angular size of 5$^{\prime\prime}$.4 $\times$ 3$^{\prime\prime}$.0 and MM3, located east of MM1, has an angular size of 6$^{\prime\prime}$.3 $\times$ 2$^{\prime\prime}$.8. Using the methanol multi-beam (MMB) survey position of the masers, the CH$ _{3} $OH maser is found to be located $\sim$\,0\arcsec.22 from the peak continuum emission (MM1). Using Equation \ref{mass equation} we also adopted a temperature of 20 $\pm$ 5 K \citep{Wu2006, Urquhart2011, Lu2014}, since an \ion{H}{II} region has not yet been detected in G22.357+0.066 and G25.411+0.105. G25.411+0.105 is located at a distance 9.5 $\pm$ 0.9 kpc \citep{Szymczak2007} and the core mass of G25.411+0.105 (MM1) was estimated to be $ \sim $ 471$\pm$82 M$_{\odot}$. The mass for cores MM2 and MM3 were estimated to be $ \sim $ 192$\pm$42 M$_{\odot}$ and 98$\pm$16 M$_{\odot}$, respectively. 

\subsubsection{Carbon-monoxide Emission in G25.411+0.105}

 The channel maps for the CO emission is not shown for G25.411+0.105 since they all show the same compact emission as seen in the integrated intensity maps,  which are indicated by the 3-color image in Figure \ref{fig:G25_continuum}, with red as the $^{12}$CO emission, green the $^{13}$CO emission and the C$^{18}$O emission is indicated in blue.  Unlike the case of G22.357+0.066, most of the CO emissions are spatially distributed around the peak of the dust continuum core (MM1). The $^{12}$CO integrated intensity map shows CO emission associated with MM2 and MM3, west and east of the dust continuum (MM1), respectively and the $^{13}$CO and C$^{18}$O emission are mostly associated with MM1.
 
 \begin{figure}
	\includegraphics[width=\columnwidth]{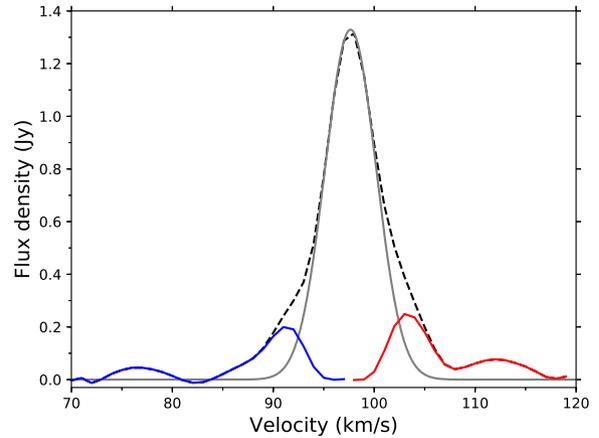}
    \caption{G25.411+0.105: $^{13}$CO line spectrum (black), extracted at 80 $\%$ flux density from the $^{13}$CO channel maps, with the Gaussian fit (grey). The blue and red lines indicate the blue-shifted and red-shifted emission components (wings).}
    \label{fig:G25_13CO_linefit}
\end{figure}

Figure \ref{fig:G25_13CO_linefit} presents the line spectrum of the $^{13}$CO emission detected towards G25.411$+$0.105. The Gaussian fit as well as the wing components are indicated on the spectrum.  The systemic velocity of G25.411+0.105 is at 96.0 km s$^{-1}$ \citep{Szymczak2007}, which was determined by the  line peak velocities of C$^{34}$S (2-1) and C$^{34}$S (3-2). The blue-shifted (85.0 km s$^{-1}$ to  95.0 km s$^{-1}$) and red-shifted (100.0 km s$^{-1}$  to 107.0 km s$^{-1}$) components (wings) are highlighted in blue and red, respectively. The wing channels were extracted as emission associated with out-flowing gas and will be discussed in section \ref{Bipolar outflows}. \citet{Szymczak2007} also found 'wings' in the $^{13}$CO spectrum of G25.411+0.105 and they estimated a velocity range of 11 km s$^{-1}$ for the $ ^{13} $CO wings.

\subsection{G9.62+0.19E} 
\subsubsection{1.3 mm Continuum Emission: G9.62+0.19E}
\label{subsec:G9.62}
\citet{Liu2017} reported 12 millimetre dust continuum cores (MM1 -- MM12) for G9.62+0.19
that was detected with ALMA. Our main focus is on the MM4 core (G9.62+0.19E)  located at
$(\alpha, \delta)_{\mathrm{J}2000.0}= (18^{\mathrm{h}}06^{\mathrm{m}}14^{\mathrm{s}}.66$,
-20$^{\circ}$31\arcmin31\arcsec.6) associated with the periodic methanol
masers. Previous observations have reported an ultra-compact H\,II region and millimetre dust
continuum toward G9.62+0.19E e.g. \citet{Sanna2015, Liu2017, Testi2000} and
\citet{Hofner1996}.

\citet{Liu2017} reported the peak intensity of the Gaussian component of the dust emission
to be 0.26$\pm$0.01 mJy\,beam$^{-1}$ and the residual point-like component peak intensity
to be 0.93$\pm$0.03 mJy\,beam$^{-1}$. They derived the effective radius of 1\arcsec.65
corresponding to 8600 au and a core mass of 74.0$\pm$23.4
M$_{\odot}$ (assuming a dust temperature of 100 K) and a density of
(4.2$\,\pm$\,1.3)$\times$10$^6$~cm$^{-3}$. \citet{Liu2017} also detected hot
molecular lines such as methyl formate (CH$_{3}$OCHO $\nu_{t} = 1 $ (17$ _{4,13} $-16$
_{4,12} $)) and strong methanol ( CH$_{3}$OH $\nu_{t} = 1 $, (6$ _{1} $-7$ _{2} $))
lines toward G9.62$+$0.19E, making it a hot core. \citet{Liu2011}  derived a rotational temperature of
92 K using six transitions thioformaldehyde ($\mathrm{H _{2}CS}$).

\begin{figure}
	\includegraphics[width=\columnwidth]{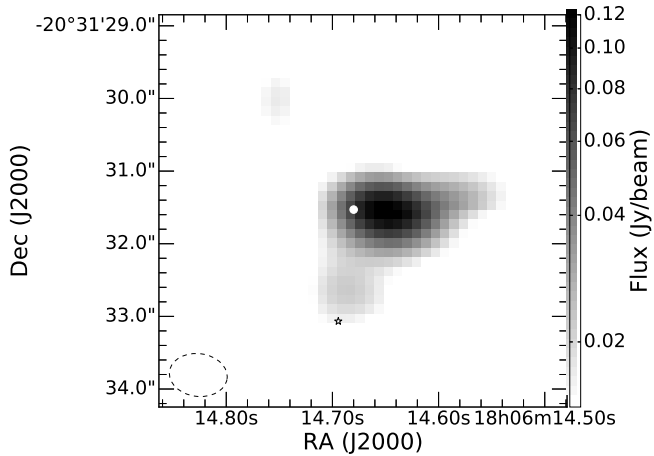}
    \caption{ G9.62+0.19E: 1.3 $mm$ continuum emission with the NIR counterpart (star
      symbol) obtained from \citet{UKIDSS2012} within a radius of 2 $ ^{\prime \prime} $
      from the maser. The circle shows the position of the periodic 6.7 GHz methanol
      maser with a $\sim$ 6$^{\prime\prime}$ uncertainty. The synthesized beam is shown in the left bottom as an ellipse.}
    \label{fig:G9_continuum}
\end{figure}

Figure \ref{fig:G9_continuum} presents the millimetre dust continuum emission of
G9.62+0.19E. The methanol maser is marked with a circle and the star symbol indicates
the NIR counterpart obtained from the database of the UKIRT Infrared Deep Sky Survey
(UKIDSS).  Table \ref{tab:cont parameters} presents the peak position (RA and Dec) and
other continuum parameters G9.62+0.19E (MM4).  G9.62+0.19E is located at a distance of 5.2 $\pm$ 0.6 kpc \citep{Sanna2009}. Thus from Equation \ref{mass equation} a core
mass of $\sim$ 59$\pm$14 M$_{\odot}$ was estimated for G9.62+0.19E, adopting a dust temperature of 100 $\pm$ 5 K, since G9.62+0.19E is a hot core with a \ion{H}{II} region \citep{Lu2014} and \citet{Liu2017} derived a rotational temperature of 92 K for $\mathrm{H _{2}CS}$. The uncertainties in the dust temperature and the distance causes the main uncertainty in the core mass.

\subsubsection{ $^{12}$CO Emission in G9.62+0.19E}

\begin{figure*}
	\includegraphics[ width=0.7\textwidth]{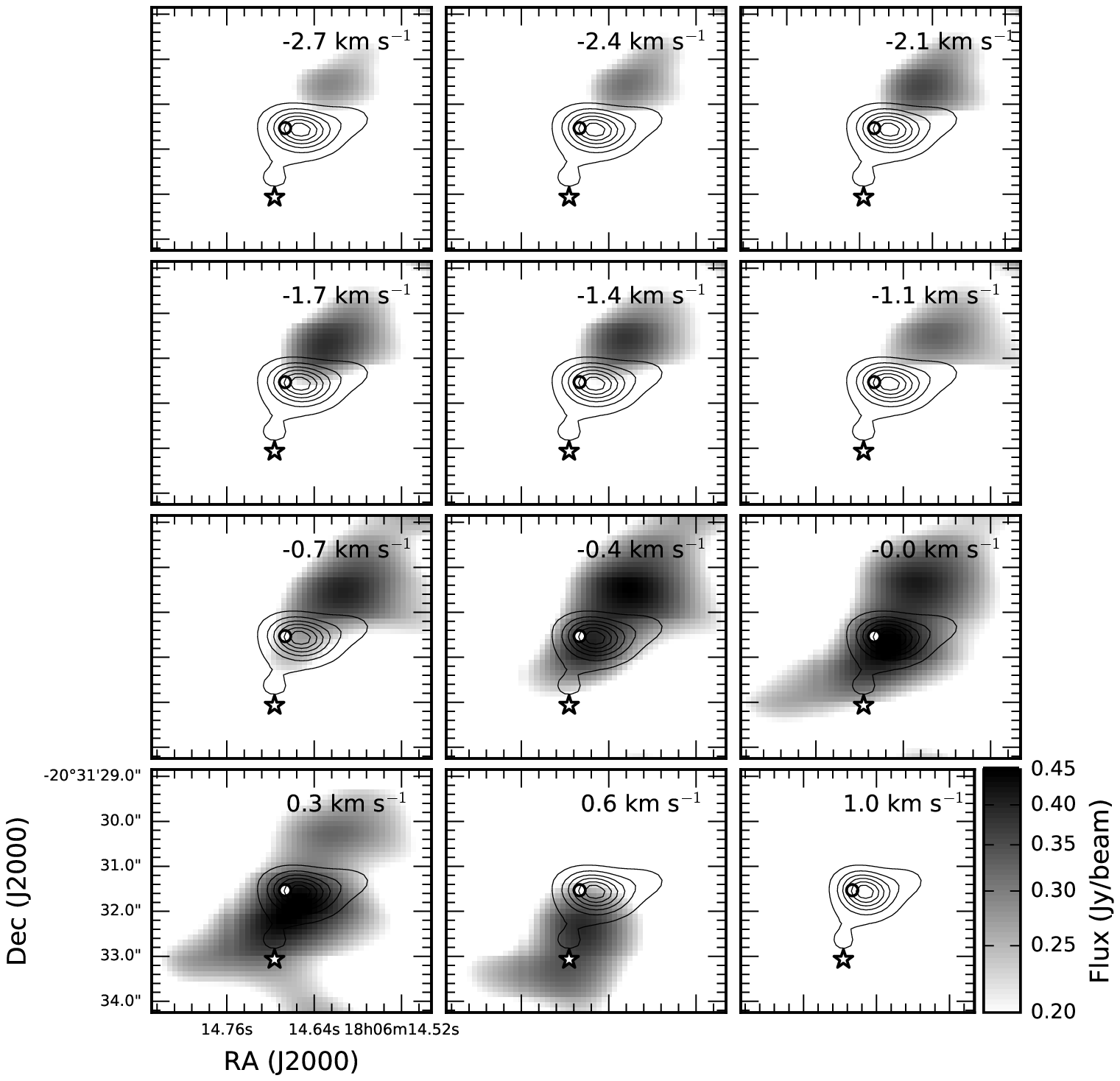}
    \\[\smallskipamount]
	\includegraphics[width=0.7\textwidth ]{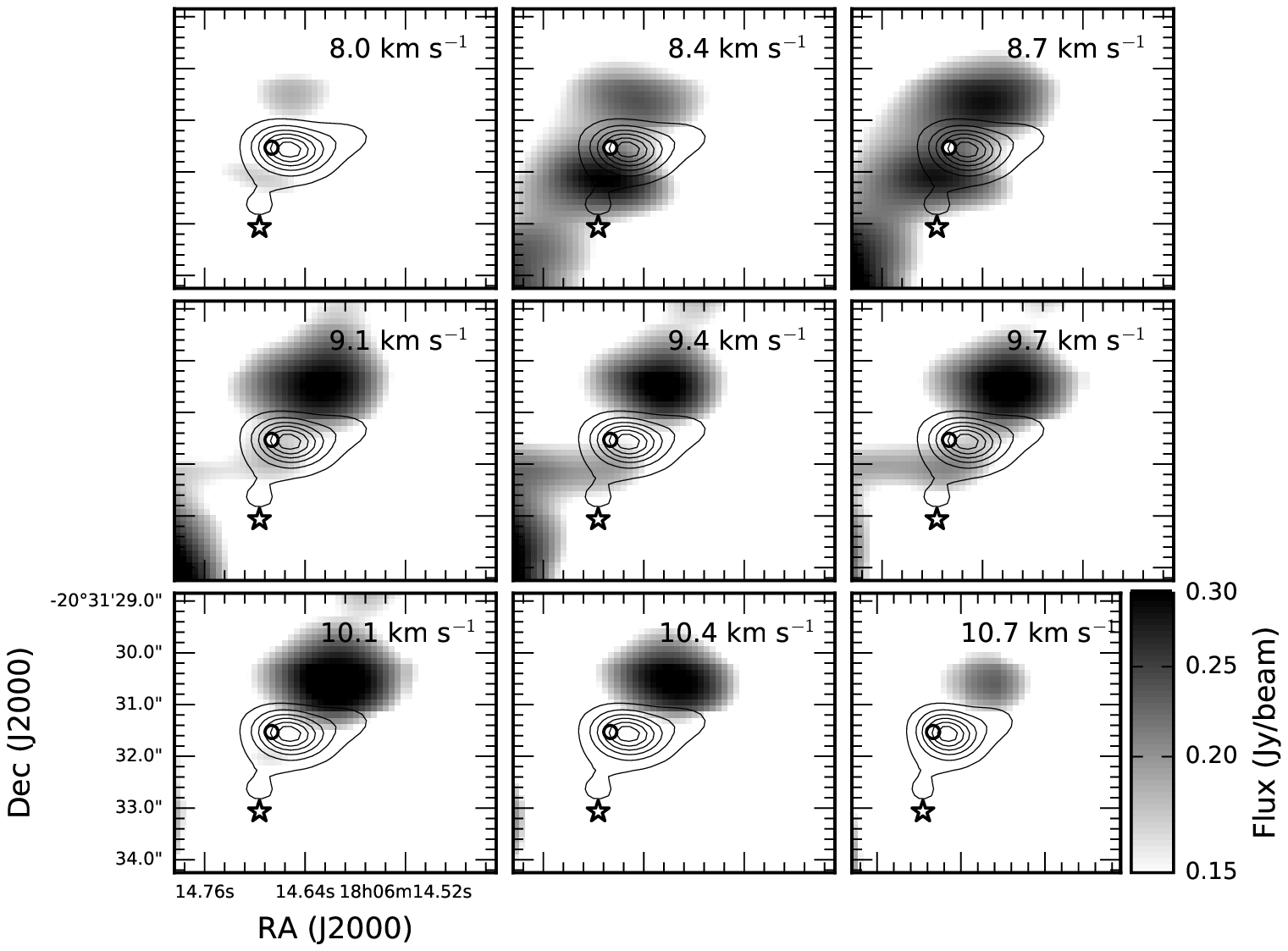}
    \caption{ G9.62+0.19E: $^{12}$CO blue-shifted velocity channel maps from -2.7 km s$
      ^{-1} $ to 1.0 km s$ ^{-1} $ and the red-shifted velocity channel maps from 8.0 km s$
      ^{-1} $ to 10.7 km s$ ^{-1} $.  The position of the periodic methanol maser is
      indicated with a circle and the dust continuum contours levels are [20, 40, 60, 80, 100, 120, 140] $\times$ $\sigma$ mJy\,beam$^{-1}$, where $\sigma$ is 1mJy\,beam$^{-1}$. The NIR counterpart obtained from the UKIDSS database is indicated with the star symbol.}
    \label{fig:G9_12CO_chanmaps}
\end{figure*}

Unlike our SMA observations, the archival ALMA data of G9.62+0.19E only covers the
$^{12}$CO transition and not the $^{13}$CO and C$^{18}$O lines. Figure \ref{fig:G9_12CO_chanmaps} 
show the $^{12}$CO channel maps, overlaid is the dust continuum emission (black contours)
and the position of the periodic 6.7 GHz methanol maser is indicated by a circle.  The
systemic velocity of G9.62+0.19E is 2.1 km s$ ^{-1} $ \citep{Liu2017} determined from the line peak velocity of  CH$_{3}$OH $\nu_{t}=1, $ (6$ _{1} $-7$ _{2} $), which makes
channel maps -2.7 km s$ ^{-1} $ to 1.0 km s$ ^{-1} $ blue-shifted emission and and channel maps 8.0 km s$ ^{-1} $ to 10.7 km s$ ^{-1} $ red-shifted emission. Figure \ref{fig:G9_12CO_chanmaps}  shows an emission core north-west of the dust continuum at  -2.7 km s$ ^{-1} $, the core peaks at  -0.4 km s$ ^{-1} $ and disappears after 0.3 km s$ ^{-1} $. A second blue-shifted emission core appears at -0.4 km s$ ^{-1} $ on the dust continuum and extend to the south-east of the dust continuum, after which it disappears at 0.6 km s$ ^{-1} $. Figure  \ref{fig:G9_12CO_chanmaps}  shows two red-shifted emission cores that appear at 8.0 km s$ ^{-1} $. One core is located north-west of the dust continuum and the other is south-east of the dust continuum. The south-east core disappears after 9.7 km s$ ^{-1} $ and the north-west core peaks at 10.1 km s$ ^{-1} $, after which it disappears at 10.7 km s$ ^{-1} $. Thus there are two blue-shifted emission cores and two red-shifted emission cores found in the $^{12}$CO emission, which indicates to evidence of a north-west -
south-east multi-bipolar outflow system and will be discussed in section \ref{Bipolar outflows}.

\begin{figure}
\includegraphics[width=\columnwidth]{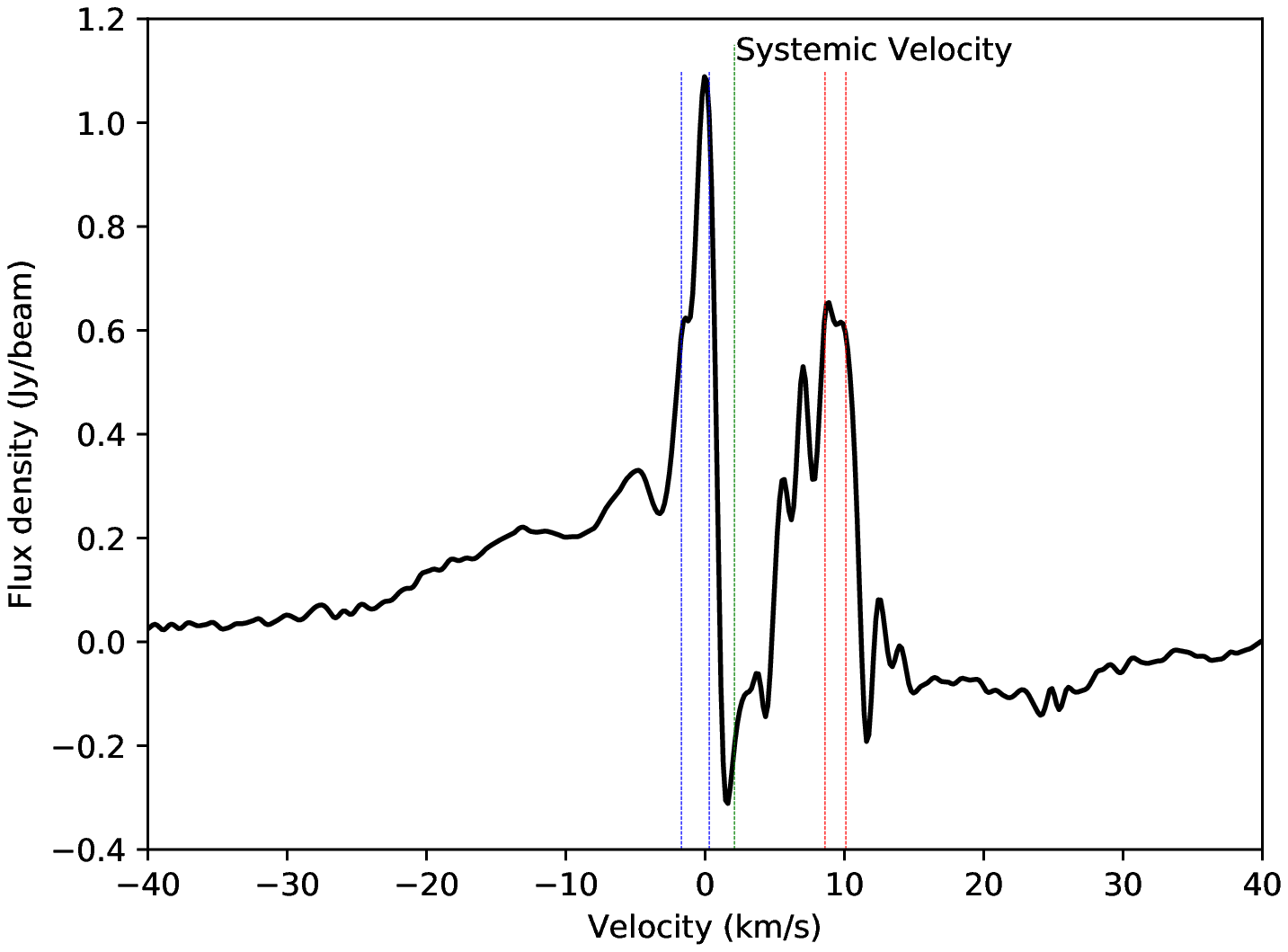}
    \caption{$^{12}$CO spectrum of G9.62$+$0.19E. The systemic velocity (2.1 km s$ ^{-1} $) is indicated in green and the peak intensities of the blue- and red-shifted cores are indicated in blue (-1.7 km s$ ^{-1} $ and 0.3 km s$ ^{-1} $) and red (8.3 km s$ ^{-1} $ and 10.1 km s$ ^{-1} $), respectively}
    \label{fig:G9_12CO_spectrum}
\end{figure}

\subsubsection{Rotation in the methanol of G9.62+0.19E \label{Rotation in the methanol} }
\begin{figure*}
	\includegraphics[width=15cm]{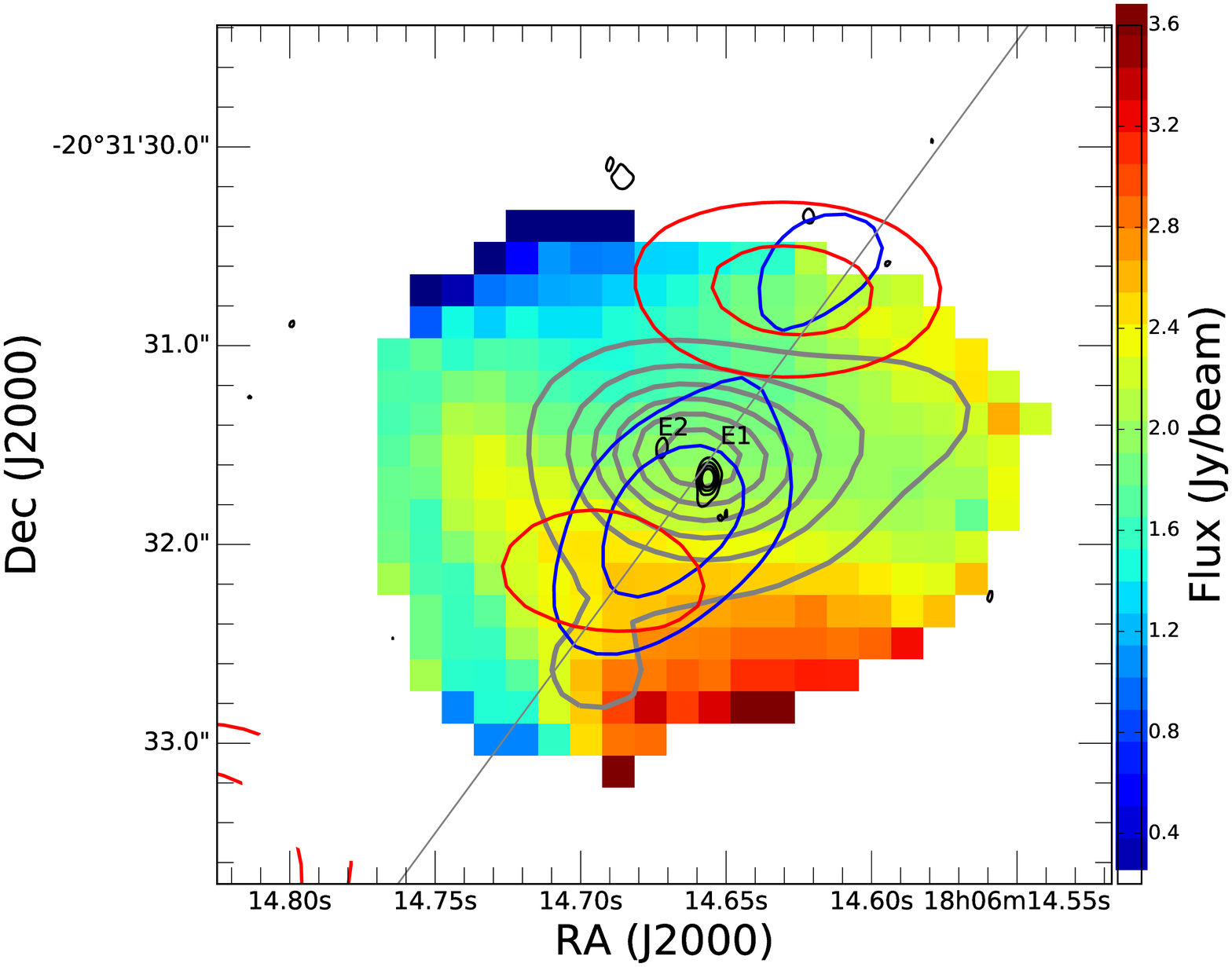}
    \caption{ Moment 1 map of CH$_{3}$OH  $v_{t}=1$, (6$_{1}$-7$_{2}$ A-) emission of
      G9.62+0.19E obtained from \citet{Liu2017}. The dark grey solid contour lines is the dust continuum and has the same levels as indicated in Figure \ref{fig:G9_continuum}. The blue and red contours indicate the peak intensities of the blue- and red-shifted emission as seen in the $^{12}$CO line spectrum (Figure \ref{fig:G9_12CO_spectrum}). E1 and E2 represents the sub-continuum cores detected by \citet{Sanna2015}  with the contour levels [1.0, 2.0, 3.0, 4.0] mJy\,beam$^{-1}$. The grey line indicates the direction of the bipolar outflow in G9.62$+$0.19E.}
    \label{fig:G9_CH3OH_mommap1}
\end{figure*}

Figure \ref{fig:G9_CH3OH_mommap1} shows the  CH$_{3}$OH $\nu_{t}=1 $, (6$ _{1} $-7$ _{2} $) detected towards G9.62+0.19E. The CH$_{3}$OH emission has a compact morphology on the dust continuum and \citet{Liu2017} fitted the CH$_{3}$OH spectrum with a Gaussian profile and obtained a FWHM of 4.70 $\pm$ 0.02 km s$^{-1}$. The methanol data was explored further in search for rotation, since it might lead to a
better understanding of the multi-outflow system observed in the $^{12}$CO velocity
channel maps. Figure \ref{fig:G9_CH3OH_mommap1} is the velocity field map (moment 1 map)
of CH$_{3}$OH  $v_{t}=1$, (6$_{1}$-7$_{2}$ A-) obtained from the data of \citet{Liu2017} and  shows a
spatial separation of the red- and blue-shifted emission, indicating evidence of
rotational motion with the enclosed mass being $\sim$ 70 \% of the core mass. These velocity gradients in the methanol moment 1 map that suggests
rotation were also reported by \citet{Liu2017}. In addition, \citet{Liu2011} (their Fig. 4) also reported rotation associated with G9.62+0.19E using the H$ _{2}$CS ($10_{2,8}-9_{2,7}$) line.  The red and blue contours superimposed on the methanol were obtained form the peak intensities of the blue- and red-shifted emission in the 12CO (Figure \ref{fig:G9_12CO_spectrum} and the line indicates the general orientation of the bipolar outflows). VLBI observations toward G9.62+0.19E by \citet{Sanna2015} revealed two continuum cores within core E and are indicated as sub-continuum cores E1 and E2.

\section{Discussion}
\subsection{Disk-outflow orientation \label{Bipolar outflows} }

\begin{figure}
	\includegraphics[width=\columnwidth]{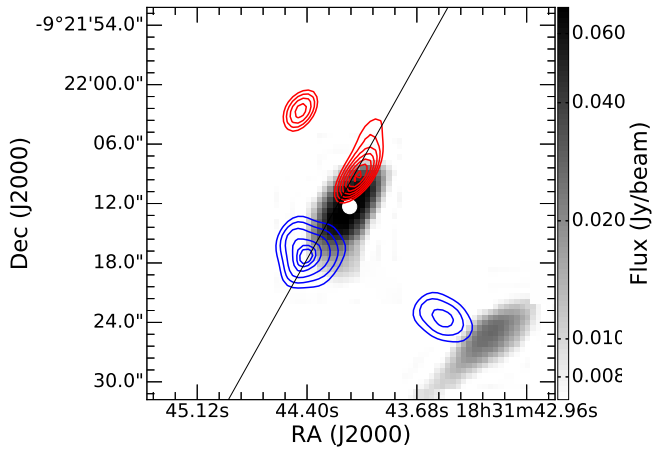}
    \caption{ Bipolar outflow of G22.357+0.066 traced by the $^{12}$CO emission. The red and blue contours that indicate the blue-shifted (72 km s$^{-1}$ $\sim$ 80 km s$^{-1}$) and red-shifted (88 km s$^{-1}$ $\sim$ 92 km s$^{-1}$) emission respectively are superimposed on the dust continuum. The black line indicates the the direction of the bipolar outflow. The circle is the periodic methanol maser.}
    \label{fig:G22_12CO_outflows}
\end{figure}

\begin{figure}
    \includegraphics[width=\columnwidth]{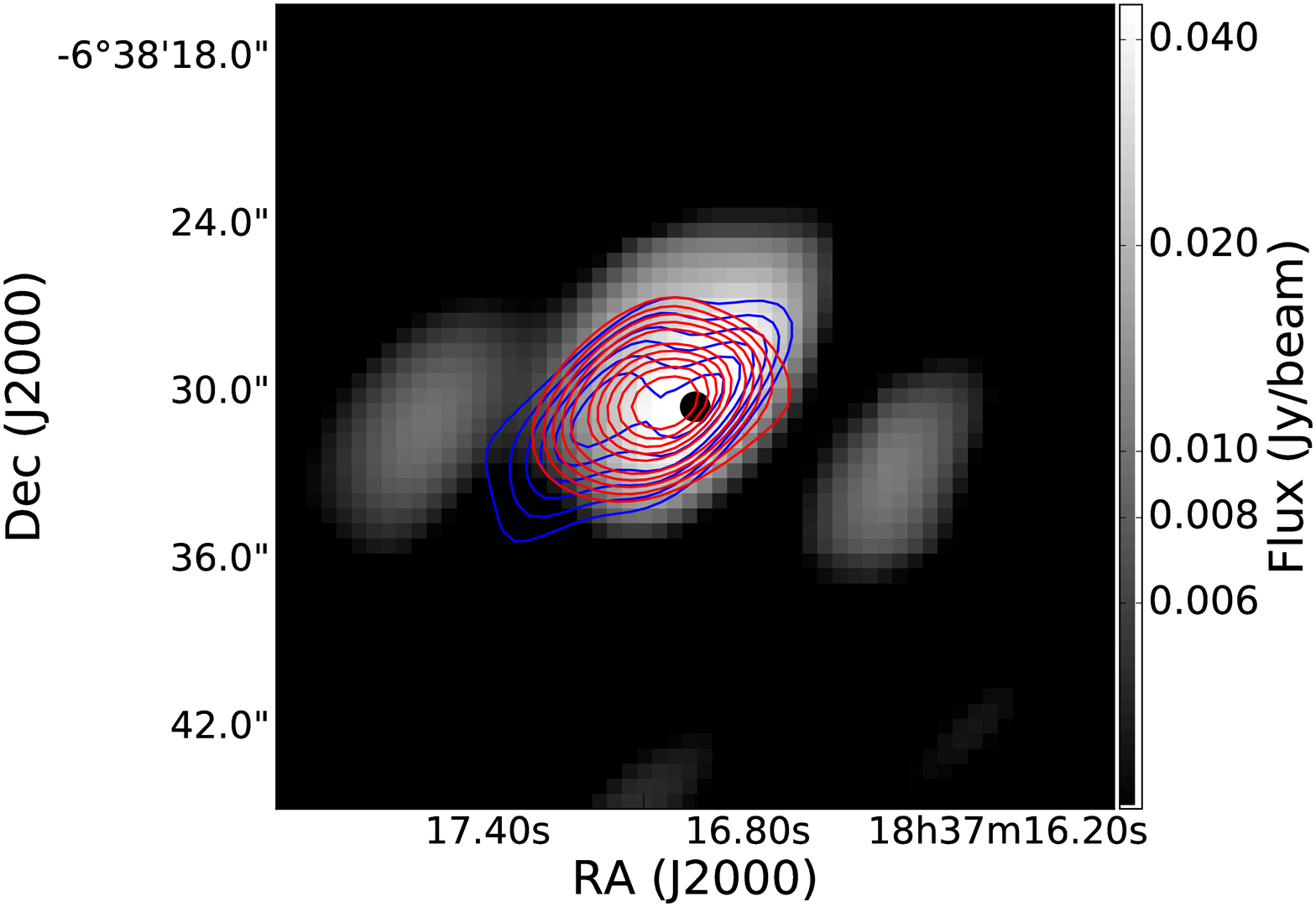}
     \caption{ Bipolar outflow of G25.411+0.105 traced by the $^{13}$CO emission. The red
      and blue contours indicate the blue-shifted (89 km s$^{-1}$ $\sim$ 91 km s$^{-1}$) and red-shifted (102 km s$^{-1}$ $\sim$ 104 km s$^{-1}$) emission respectively that are superimposed on the dust continuum. The circle is the periodic methanol maser.}
    \label{fig:G25_13CO_outflows}
\end{figure}

The $^{12}$CO velocity channel maps of G9.62+0.19E (Figure \ref{fig:G9_12CO_chanmaps}) show a clear north-west - south-east multi-outflow structure. The case of G9.62+0.19E is not as simple and requires further consideration. Although \citet{Sanna2015} identified a second sub-continuum source (E2) offset in projection by 1300 AU from sub-continuum core E1 in G9.62+0.19E, the periodic masers are clearly associated with core E1 as is seen from Figure 1 of \citet{Sanna2015}. This applies also to the periodic OH masers associated with this region as argued by \citet{Goedhart2019}. The exact relation between components E1 and E2 is not clear even though both of these are projected inside the core as is evidenced in Figure \ref{fig:G9_CH3OH_mommap1}. The axis of the outflow suggest that it is driven by the YSO associated with E1. However, the angular resolution of our data is not sufficient to confirm whether some of the outflowing gas emanate from E2. Considering the velocity field map (Figure \ref{fig:G9_CH3OH_mommap1}) of
G9.62+0.19E, the velocity gradient seen in the CH$_{3}$OH $v_{t}=1$,
(6$_{1}$-7$_{2}$ A-) points to a disk with an inclination along the line of sight.

The symmetry (south-east and north-west emission) seen in the $^{12}$CO channel maps (Figure \ref{fig:G22_12CO_chanmaps}) of G22.357+0.066 shows evidence of outflow from a central core (MM1). The emission south-east and north-west of G22.357+0.066-MM1 was extracted and imaged as red- and blue-shifted emission and is presented in Figure \ref{fig:G22_12CO_outflows} as a bipolar outflow. The $^{12}$CO emission in G22.357+0.066 is quite complex, but channels 79 km s$^{-1}$ to 80 km s$^{-1}$ and 89 km s$^{-1}$ to 90 km s$^{-1}$ of $^{12}$CO (Figure \ref{fig:G22_12CO_chanmaps}), indicate bipolarity in their symmetry which cut through the peak position of G22.357+0.066-MM1

The $^{13}$CO spectrum (Figure \ref{fig:G25_13CO_linefit}) of G25.411+0.105 shows wings, which is interpreted as outflow from a central core. The $^{13}$CO channels falling in the velocity ranges of the blue and red-shifted components were extracted and imaged as blue- and red-shifted emission. Figure \ref{fig:G25_13CO_outflows} present the bipolar outflow of G25.411+0.105. The red- and blue-shifted components of the outflow are spatially coincident, which indicates that the outflow direction is along the line of sight of the observer. Assuming a disk-outflow system, then the rotating structure around G25.411+0.105 must be face-on with the axis of rotation closely aligned with the line of sight.

{\ The inclination angle (relative to the line-of-sight) in G9.62+0.19E and G22.357+0.066 was calculated using the method given in \citet{Kong2011}. For G22.357+0.066 the maser position was chosen as the center of the driving source and a inclination angle of $\sim 82^{\circ}$ was obtained. For G9.62+0.19E two driving source positions were chosen, the maser and core E1. The inclination angle obtained using the maser is $\sim 82^{\circ}$ and the inclination angle found using E1 as the position is the driving source is $\sim 57^{\circ}$. The orientation (inclination angles) of the bipolar outflow of G9.62+0.19E ($\sim 71^{\circ}$) and G22.357+0.066 $\sim 82^{\circ}$ are similar to
one another, but different from the outflow orientation of G25.411+0.105.} In the case of
G22.357+0.066 and G9.62+0.19E the bipolar outflows are  almost in the plane (inclination of $\sim 82 ^{\circ}$ and $\sim 71 ^{\circ}$, respectively) of the sky. The method used in \citet{Kong2011} can not be applied to G25.411+0.105 to obtain a inclination angle, since it can only be used for a conical-like outflow. The presence of bipolar outflows also imply the existence of disks
\citep{Zhang2007}. Thus, since G22.357+0.066 and G9.62+0.19E have outflows orientated almost in the
plane of the sky, their disks would be viewed  approximately edge-on with some inclination. For G25.411+0.105 which has an
outflow orientated almost perpendicular to the plane of the sky, the disk would then be viewed
face-on. 

\begin{figure}
{\bf G9.62$+$0.19E}\\
	\includegraphics[ width=1.0\columnwidth]{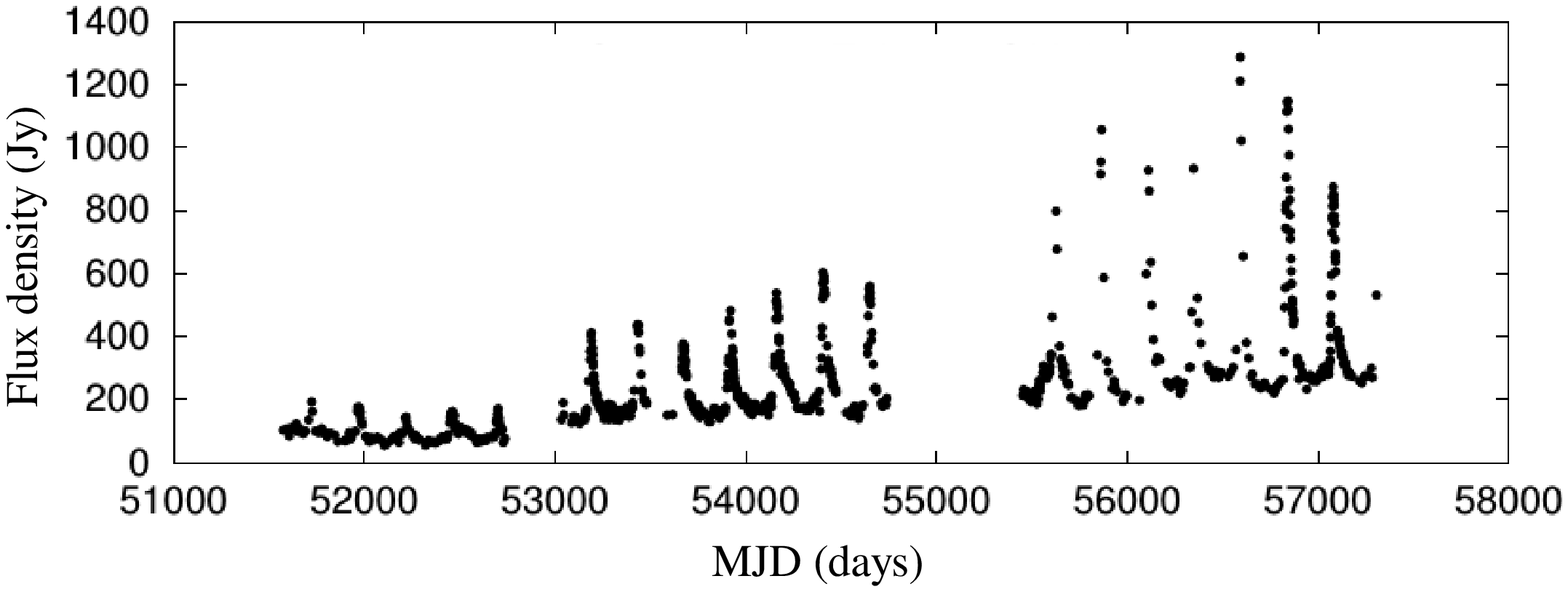}
    \\[\smallskipamount]
{\bf G22.357$+$0.066} \\   
	\includegraphics[ width=1.0\columnwidth]{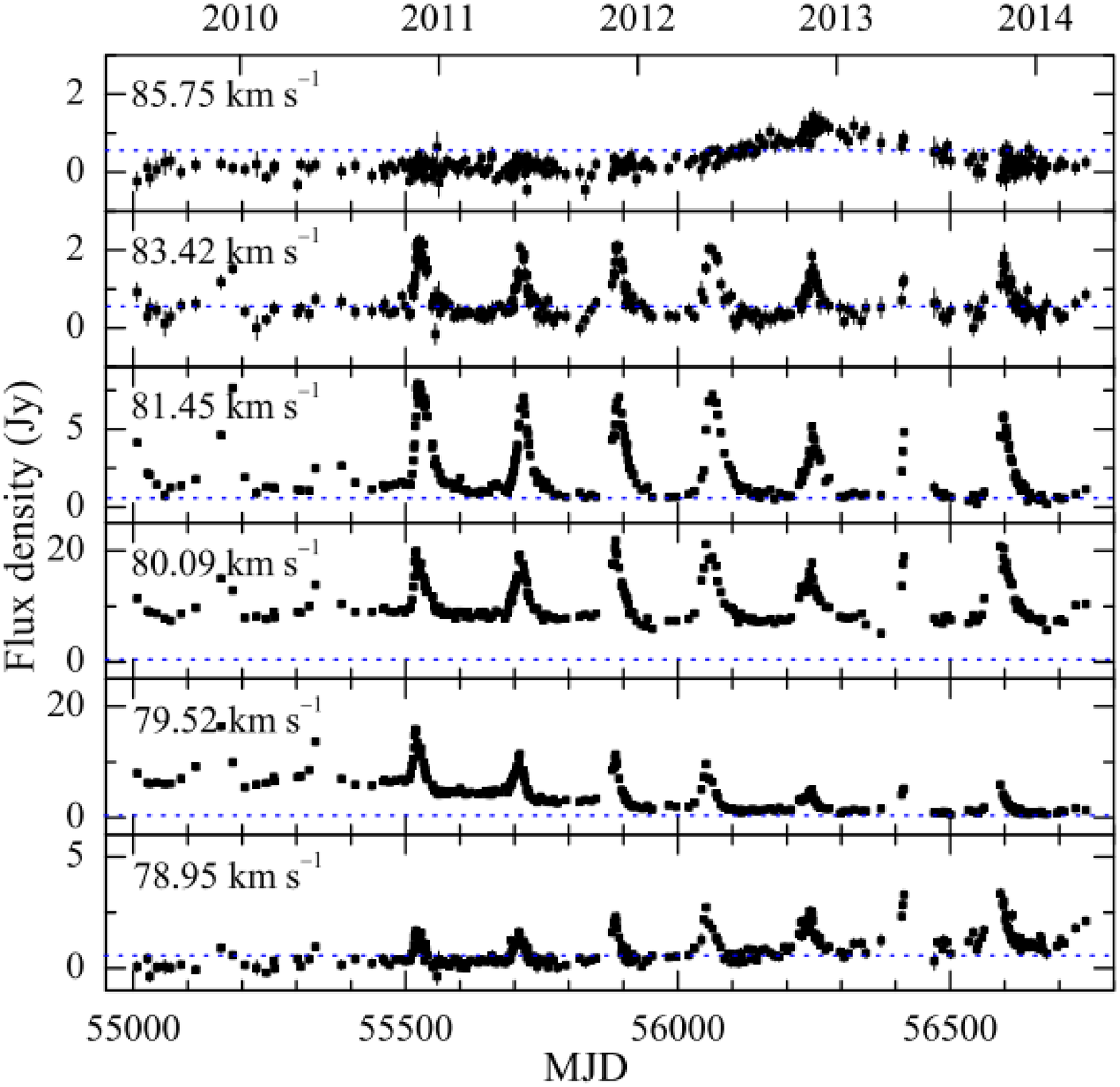}
    \\[\smallskipamount]
{\bf G25.411$+$0.105} \\   
	\includegraphics[width=1.0\columnwidth]{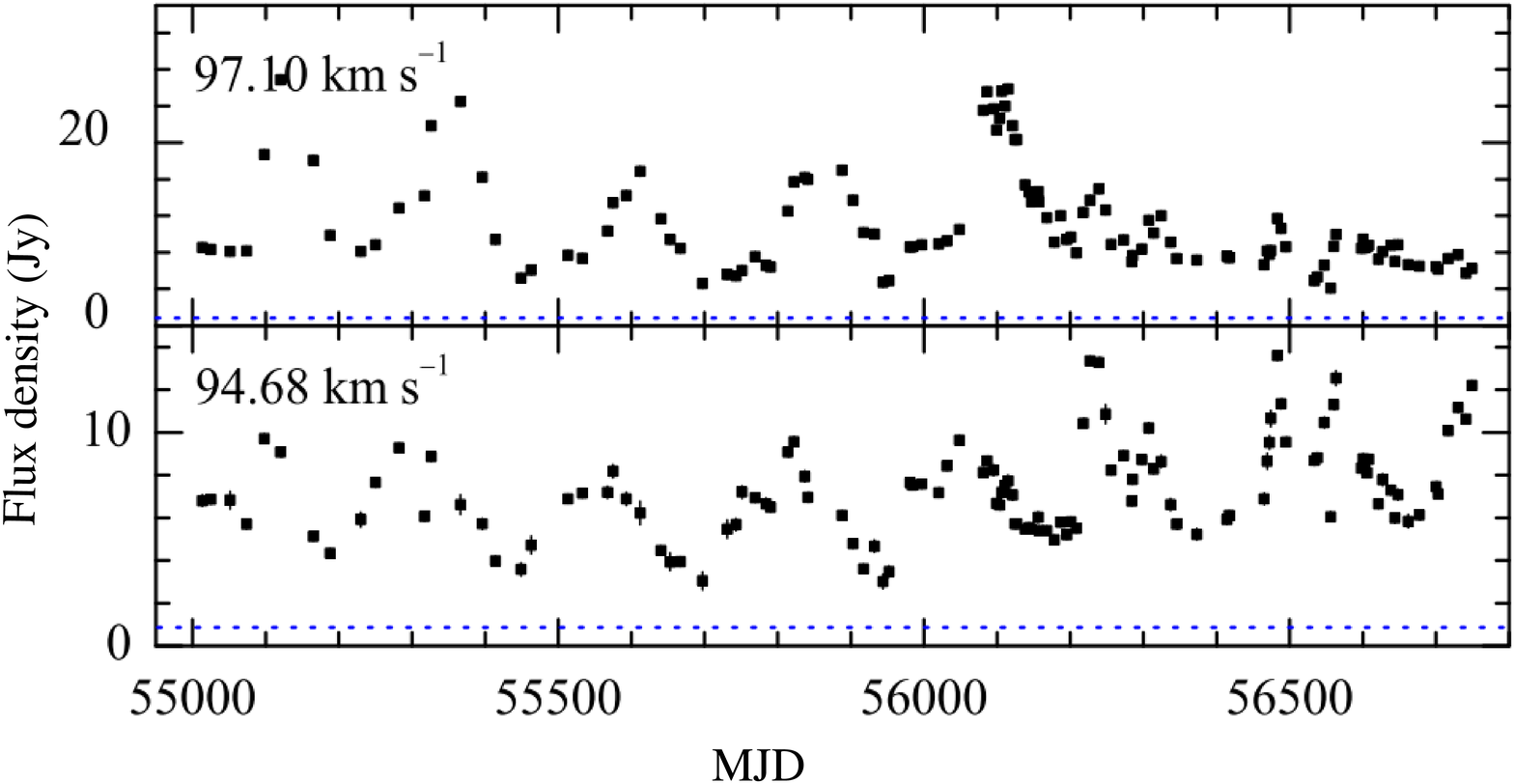}
    \caption{Light curve profiles of G9.62$+$0.19E obtained from \citep{Goedhart2014}, G22.357$+$0.066 and G25.411$+$0.105, both obtained from \citep{Szymczak2015}.}
    \label{fig:light_curves}
\end{figure}

From Figure \ref{fig:light_curves}, it can be seen that the periodic methanol masers of G9.62$+$0.19E and G22.357$+$0.066 have similar light curves and they also have similar outflow orientations (Figures \ref{fig:G9_CH3OH_mommap1} and \ref{fig:G22_12CO_outflows}). Based on the presence of an {H\,{\scriptsize II}} region in G9.62+0.19E and the non-detection thereof in
G22.357+0.066 by \citep{Bartkiewicz2011}, G9.62+0.19E is likely to be more evolved
than G22.357+0.066. Figure \ref{fig:light_curves} also shows that G25.411+0.105 has a different periodic  methanol maser light curve from that of G9.62+0.19E and G22.357+0.066 and has a different outflow orientation (Figure \ref{fig:G25_13CO_outflows}). This may point to inclination and/or orientation effects as a possible cause of the difference in the flare profiles of the periodic methanol maser considered here.

\subsection{Possible Explanations of the Impact of Source Orientation}
\begin{figure}
  \includegraphics[width=\columnwidth]{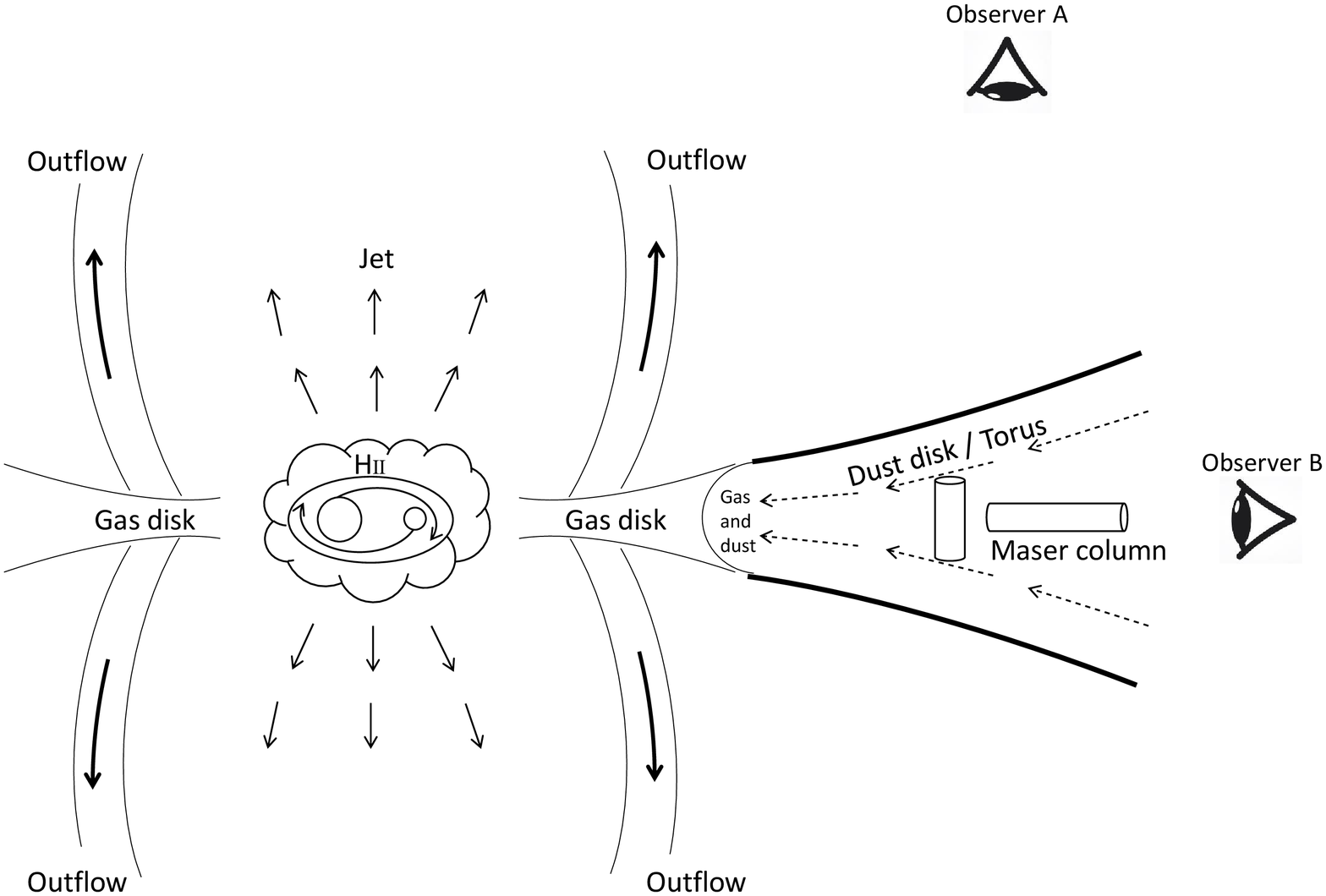}
  \caption{Schematic to illustrate the effect of different orientations of the masing columns of gas. 
  Adapted from \citet{Vaidya2009}.}
  
  \label{fig:schematic}
\end{figure}

Although the results presented above are for only three sources, it nevertheless is rather    remarkable that for G9.62+0.19E and G22.357$+$0.066, which have similar light curves and which can be explained within the framework of the CWB scenario, the observations suggest that the outflows associated with each of the sources are projected almost in the plane of the sky. On the other
hand, G25.411+0.105, with a completely different periodic maser light curve (bunny hop), seems to have an orientation such that the associated outflow is almost perpendicular to the plane of the sky. The question is how and whether such differences in the orientation of the source can play a role in determining
the shape of the light curve of the periodic masers.

It is not yet clear where in the environment of the young high-mass stars the methanol
masers arise. However, \citet{Vanderwalt2016} argued that the shape of the maser light
curve reflects the underlying mechanism that gives rise to the periodicity. Implicitly
this means that the masing region's position and orientation must be such that the
variation in the maser emission reflects the effects of the underlying mechanism of the
periodicity. For G9.62+0.19E and G22.357$+$0.066 the implication within the CWB scenario is that the periodic masers must be projected against an \ion{H}{II} region. Given the observational results for the three periodic masers, we argue here that the difference in the light curves of G9.62+0.19E and G22.357$+$0.066 on the one hand and G25.411+0.105 on the other hand may be due to the masers in these three sources being located in accretion disks or rotating toroidal structures but have different
orientations in these structures. By making this assumption we do not exclude other
possibilities for where the masers may arise (eg. from outflowing gas) but apply it only to the current discussion.

The scenario we propose is a combination of an inner disk as that shown in Fig. 6 of
\citet{Vaidya2009} and a rotating torus as in Fig. 9 of \citet{Sollins2005b} with the
inner disk being feeded with material from the rotating torus. Comparison of
Fig.\,\ref{fig:G9_CH3OH_mommap1} with the equivalent maps of G10.62-0.38, G19.61-0.23 and
G29.96-0.02 \citep{Beltran2011} suggests that a rotating toroid is also present in
G9.62+0.19E. Associated with the disk-torus system is a bipolar outflow which carved out a
conical shaped cavity in the central part of the torus. Depending on the evolutionary
state of the high-mass star, a small \ion{H}{II} region may be present similar to that
proposed by \citet{Sollins2005b} for G28.20-0.05. The \ion{H}{II} region may be a trapped
\ion{H}{II} region \citep{Keto2002} or it may already have expanded into the lower density
cavity created by the outflow as suggested in Fig. 9 of \citet{Sartorio2019}.  For the
masers we consider two cases for the orientation of the masing columns, ie. perpendicular
to the plane of the disk-torus system or in the plane of the disk-torus system. The assumption that 
the masers are located in a disk is not new. \citet{Parfenov2014} made a similar assumption 
in their model. A schematic representation of this scenario is shown in Fig.\,\ref{fig:schematic}.

Consider first the edge-on view as seen from the vantage point of observer B as shown in Fig.\,\ref{fig:schematic}. It is quite obvious that masers in the plane of the disk/torus can indeed be projected against the \ion{H}{II} region and that the CWB scenario is in agreement with such a geometry. Given that the maser flare profiles of G9.62+0.19E and G22.357$+$0.066 can be explained within the CWB scenario, and that the observed outflows for these two sources are projected almost in the plane of the sky, suggest that the geometry as for observer B in Fig.\,\ref{fig:schematic} might apply to these two sources.  Depending on the physical conditions in the  disk/torus, it is possible that some masers in the plane of the disk/torus might not be projected directly against the \ion{H}{II} region. However, the possibility of not being projected against the \ion{H}{II} region is not as severe as in the case for masers oriented perpendicular to the plane of the disk/torus as discussed next.

Thus, consider the case in which the disk-torus is viewed face-on, ie. from the position
of observer A in Fig.\,\ref{fig:schematic}. This scenario will correspond to that of
G25.411+0.105, viz. where the observed outflow is almost perpendicular to the plane of the sky. The effect of the outflow and the cavity it creates is that masers located
in the disk and which are pointing towards the observer cannot be projected against the
\ion{H}{II} region. Also, due to the radial dependence of the density and temperature in
the disk/torus it is reasonable to assume that the masers will arise at those radial
positions where the conditions for pumping are optimal. According to \citet{Cragg2002} the
conditions under which class II methanol masers are pumped are for $n_{H_2} <
10^8~\mathrm{cm^{-3}}$, and $T_k < T_d \simeq 175$K with $T_k$ the gas kinetic temperature
and $T_d$ the dust temperature. \citet{Parfenov2014} set the upper limit for the density
at $n_{H_2} = 10^9~\mathrm{cm^{-3}}$ and a dust temperature $>$ 100 K.  Using these
numbers with the theoretical results of \citet{Kuiper2013}, it is found that the masers can
occur at distances greater than about 400 AU from the star provided that $T_d > 100$K. 
There is therefor at least two physically justifiable reasons why in the case of
this geometry the masers cannot be projected against the \ion{H}{II} region if they are
located in the disk/torus.

From these considerations it follows that the orientation of the disk/torus-outflow system
in the sky does play a role in ``selecting'' which scenarios for the periodic variability
of methanol masers are compatible with a particular flare profile. For example, for the CWB
scenario \citep{Vanderwalt2011, Vandenheever2019}, in which the decay of the maser follows that of a
recombining partially ionized hydrogen plasma, it is required that the masers be projected
against some part of the \ion{H}{II} region and is therefor not compatible with a face-on
orientation of the disk/torus. If in fact the face-on case implies that the periodic
masers are not projected against an \ion{H}{II} region, it follows that the underlying
cause for the periodicity is most likely related to a process that periodically changes
the physical conditions in the masing region or in it's immediate vicinity. It might,
therefore, be that in this case the proposed models of eg. \citet{Araya2010} and
\citet{Parfenov2014} are applicable. It should be noted that in the models of \citet{Araya2010} and \citet{Parfenov2014} the proposed binary system might also
be a CWB but its effect on the masers will definitely be different than for the case when
the masers are projected against the \ion{H}{II} region and the variability is not due to
    changes in the free-free seed photon flux.

In the case of G25.411+0.105, for which an outflow almost along the line-of-sight has been
detected, a ring-like distribution of masers has been mapped by
\citet{Bartkiewicz2009}. These authors fitted an ellipse to the distribution of masers,
suggesting that the disk is not viewed perfectly face-on which is in agreement with the
observation that the outflow axis is somewhat inclined relative to the line-of-sight. At a
distance of 9.5 kpc the linear radius of the ring distribution is 980 AU
\citep{Bartkiewicz2009} which agrees with the above rough estimate that the conditions for
class II methanol masers occur at radial distances greater than 400
AU. \citet{Szymczak2015} presented the time series of the maser features at 94.6 km\,s$
^{-1} $ and 97.1 km\,s$ ^{-1}$ of G25.411+0.105. A comparison of the two time series
suggests that the feature at 94.6 km\,s$^{-1}$ reaches a maximum about 30 days before the
feature at 97.1 km\,s$^{-1}$.  Using the elliptical fit of \citet{Bartkiewicz2009} and the
positions of these two features, it is found that, in projection, the angle between the
two features is about 42$^\circ$. Assuming the presence of a binary system then, for an
orbital period of 245 days, such an angular separation translates to a delay of 28.6 days,
which is surprisingly similar to the time lag in the observed time series of these two
maser features. Although it is not possible to explain the shape of the light curves for
the periodic masers, it would seem that there is real evidence for a binary system viewed
face-on.

We also searched the literature for other imaging observations done on periodic methanol maser
sources with light curves similar to that of G9.62+0.19E or those with bunny-hop type light
curves. \citet{Li2016} mapped a $^{12}$CO(1-0) outflow associated with the periodic methanol
maser source G45.473+0.133 which have a maser light curve similar to that of G9.62+0.19E. The
period of the maser is 195.7 days \citep{Szymczak2015}. The red- and blue-shifted lobes are
clearly separated even with a beam of 52$^{\prime\prime}$. Although it is not possible to
determine the exact orientation, these observations show that the outflow axis is most likely
closer to the plane of the sky than perpendicular to it. More detailed observations of this
source is required to come to a conclusion about the orientation of the outflow axis. It
nevertheless seems as if there is evidence that G45.473+0.133 shares the same properties as
G9.62+0.19E and G22.357$+$0.066, ie. a periodic maser light curve that resembles that of
G9.62+0.19E with an outflow perpendicular or nearly perpendicular to the line of sight. 

One consequence of the above described scenario is that in the case of the masers located
in a  disk/torus being viewed face-on, the maser velocities must be close to the systemic
velocity and/or have a rather small velocity range. Apart from G25.411+0.105 there are two more periodic maser sources for which the light curves has the bunny-hop behaviour,
ie. G338.93-0.06 \citep{Goedhart2014} and G358.460-0.391 \citep{Maswanganye2015}. In the
case of G25.411+0.105, the systemic velocity from the Gaussian fit in Fig. \ref{fig:G25_13CO_linefit} is 97.5 $\mathrm{km\,s^{-1}}$. \citet{Szymczak2015} report two
maser features respectively at 94.68 and 97.1 $\mathrm{km\,s^{-1}}$. Both maser features
lie within about 2.8 $\mathrm{km\,s^{-1}}$ from the systemic velocity. Using the
systemic velocity of 96.0 $\mathrm{km\,s^{-1}}$ of \citet{Szymczak2007}, the offset 
is at most 1.3 $\mathrm{km\,s^{-1}}$. \citet{Goedhart2014} reported
two maser features at -42.183 and -41.376 $\mathrm{km\,s^{-1}}$ for G338.93-0.06, thus a velocity range of only $\sim 0.8~\mathrm{km\,s^{-1}}$. The systemic velocity is taken as -44.3 $\mathrm{km\,s^{-1}}$ from the MALT90 survey \citep{Rathborne2016}. The offset from the systemic velocity is therefor less than 2.1 $\mathrm{km\,s^{-1}}$. In the case of G358.460$-$0.391 no thermal line emission spectrum could be found to determine the systemic velocity. However, the maser spectrum shows only two maser features that differ in velocity with less than 1 $\mathrm{km\,s^{-1}}$ \citep{Maswanganye2015}. 

For G9.62+0.19E, G22.357$+$0.066 and G45.473+0.133 we note the following. The velocity
range for the masers in G9.62+0.19E is at least 7.6 $\mathrm{km\,s^{-1}}$
\citep{Goedhart2014}, 4.47 $\mathrm{km\,s^{-1}}$ for G22.357$+$0.066 and $\sim$ 8
$\mathrm{km\,s^{-1}}$ for G45.473$+$0.133 \citep{Szymczak2015}. In the case of G22.357 all
the periodic masers are blue shifted with respect to the systemic velocity of 84.2
$\mathrm{km\,s^{-1}}$, with the largest velocity offset being 5.3 $\mathrm{km\,s^{-1}}$.
The systemic velocity of G45.473$+$0.133 of 63.1 $\mathrm{km\,s^{-1}}$ \citep{Szymczak2015} implies that the largest maser velocity offset from the systemic velocity is about 10 $\mathrm{km\,s^{-1}}$. In the case of G9.62+0.19E the systemic velocity is at 1.9 $\mathrm{km\,s^{-1}}$ with the blue shifted periodic masers being offset by up to 2.4 $\mathrm{km\,s^{-1}}$ and the red shifted masers up to 4.5 $\mathrm{km\,s^{-1}}$ from the systemic velocity. 

It is rather remarkable that the three sources which have periodic maser light curves that can be explained within the framework of a CWB scenario have similar properties. The same applies to the three sources for which the light curves have a bunny-hop shape, even though we have mapped only G25.411+0.105. 
\section{Conclusions}
We observed two periodic methanol maser sources, ie. G22.357$+$0.066 and G25.411+0.105 with the SMA and used archival ALMA data for G9.62+0.19E to investigate whether there are larger scale differences between the three high-mass star forming regions that might explain the difference in the maser light curves. The periodic masers in G9.62+0.19E and G22.357-0.066 have the same type of light curve while for G25.411+0.105 the light curve resembles a $\left|\cos(\omega t)\right|$ function, also referred to as a bunny-hop.

The observations indicated that both G9.62+0.19E and G22.357$+$0.066 have outflows which are aligned close to the plane of the sky. In the case of G25.411+0.105 an outflow aligned close to the line of sight was detected.  While no direct detection of accretion disks was possible with the current observations, the detected outflows suggest the presence of accretion disks in all three these systems. Considering the orientation of the outflows, our observations suggest that the accretion disks or rotating toroidal structures in G9.62+0.19E and G22.357$+$0.066 are viewed edge-on and  face-on in the case of G25.411+0.105. Together with the inferences from the maser spectra and systemic velocities of G45.473$+$0.133, G339.93$-$0.06, and G358.460$-$0.391, it is concluded that orientation effects may play a role in determining the characteristics of the light curves of periodic methanol masers.
\section*{Acknowledgements}
We thank Dr. Liu Tie and Dr. Alberto Sanna for providing us with the ALMA and VLA fits images of G9.62+0.19E, respectively. This paper makes use of the following ALMA data: ADS/JAO.ALMA\#2013.1.00957.S. ALMA is a partnership of ESO (representing its member states), NSF (USA) and NINS (Japan), together with NRC (Canada), MOST and ASIAA (Taiwan), and KASI (Republic of Korea), in cooperation with the Republic of Chile. The Joint ALMA Observatory is operated by ESO, AUI/NRAO and NAOJ. JvdW acknowledge support by the National Research Foundation of South Africa (NRF Grant Number 132494). JM and JOC acknowledge support by the Italian Ministry of Foreign Affairs and International Cooperation (MAECI Grant Number ZA18GR02) and the South African Department of Science and Technology's National Research Foundation (DST-NRF Grant Number 113121) as part of the ISARP RADIOSKY2020 Joint Research Scheme.
\section*{Data Availability}
The data used in producing the results in this article can be accessed from Submillimeter Array (SMA) and the Atacama Large Millimeter/submillimeter Array (ALMA). The raw data sets can be download from the SMA and ALMA archives.


\bibliographystyle{mnras}
\bibliography{ref} 








\bsp	
\label{lastpage}
\end{document}